\documentclass[aps,twocolumn,showpacs,superscriptaddress,floatfix]{revtex4}

\usepackage{dcolumn}
\usepackage{amsfonts,amsmath,amsthm,amssymb,bm,mathrsfs,bbm,amscd}
\usepackage{graphicx,color,ulem}
\usepackage{setspace}
\usepackage{times}

\begin{document}

\title{Efficient allocation of heterogeneous response times
in information spreading process}

\author{Ai-Xiang Cui}
\affiliation{Web Sciences Center, School of Computer Science and Engineering,
University of Electronic Science and Technology of China,
Chengdu 610054, People's Republic of China}

\author{Wei Wang}
\affiliation{Web Sciences Center, School of Computer Science and Engineering,
University of Electronic Science and Technology of China,
Chengdu 610054, People's Republic of China}

\author{Ming Tang}
\email{tangminghuang521@hotmail.com}
\affiliation{Web Sciences Center, School of Computer Science and Engineering,
University of Electronic Science and Technology of China,
Chengdu 610054, People's Republic of China}
\affiliation{Center for Atmospheric Remote Sensing(CARE),
Kyungpook National University, Daegu, 702-701, South Korea}

\author{Yan Fu}
\affiliation{Web Sciences Center, School of Computer Science and Engineering,
University of Electronic Science and Technology of China,
Chengdu 610054, People's Republic of China}

\author{Xiaoming Liang}
\affiliation{School of Physics and Electronic Engineering,
Jiangsu Normal University, Xuzhou 221116, China}

\author{Younghae Do}
\affiliation{Department of Mathematics,
Kyungpook National University, Daegu 702-701, South Korea}

\date{\today}

\begin{abstract}
Recently, the impacts of spatiotemporal heterogeneities of human activities on spreading dynamics have attracted extensive attention. In this paper, to study heterogeneous response times on information spreading, we focus on the susceptible-infected spreading dynamics with adjustable power-law response time distribution based on uncorrelated scale-free networks. We find that the stronger the heterogeneity of response times is, the faster the information spreading is in the early and middle stages. Following a given heterogeneity, the procedure of reducing the correlation between the response times and degrees of individuals can also accelerate the spreading dynamics in the early and middle stages. However, the dynamics in the late stage is slightly more complicated, and there is an optimal value of the full prevalence time changing with the heterogeneity of response times and the response time-degree correlation, respectively. The optimal phenomena results from the efficient allocation of heterogeneous response times.
\end{abstract}



\maketitle

{\bf The recent empirical studies suggest that human activities display the properties of spatiotemporal heterogeneities. More and more researchers have been devoting to understanding the impact of the heterogeneous patterns on spreading dynamics with the help of Monte Carlo simulations, as well as with analytical tools. Most of these studies show that the heterogeneities in time and space have important and at times drastic effects on the spreading dynamics. However, the impacts of heterogeneous response times (the time elapsed between receiving and replying to a message) have received less consideration. In real life, due to the limited time and energy, individuals with large degrees usually take a longer time to reply than those with small degrees. In view of this point, we assume that the response time of each individual is positive correlation of its degree, and study how the heterogeneous response times influence information spreading based on uncorrelated scale-free networks. The simulation results show that enhancing the heterogeneity of response times can facilitate the spreading dynamics in the early and middle stages, which is consistent with the theoretical analysis. Furthermore, we investigate the impact of the correlation between the response times and degrees of individuals on the spreading speed and find that information spreading can be also accelerated in the early and middle stages by reducing the correlation. But in the late stage, the full prevalence time doesn't monotonously change with the buildup of the response-time heterogeneity and the reduce of the response time-degree correlation, and there is an optimal value existed respectively. The optimal phenomena results from the reasonable allocation of heterogeneous response times. This work provides us further understanding and new perspective in the impact of human activities' spatiotemporal heterogeneities on information spreading.}

\section{\label{sec:level1}Introduction}
The quantitative understanding of the impacts of human behaviors on epidemic-like spreading dynamics has attracted widespread attention in recent years~\cite{boccaletti2006complex,bansal2010dynamic,holme2011temporal,funk2010modelling,tang2009epidemic,zhao2012epidemic}, such as the structures of contact networks, the temporal and spatial patterns of human activities. Previous researches have shown that underlying network structures can strongly affect spreading dynamics through features such as scale-free degree distribution~\cite{pastor2001epidemic,tang2009influence}, community structure~\cite{liu2005epidemic,gong2012variability}, and degree correlation~\cite{boguna2003epidemic}. Most of these researches have neglected human temporal activity patterns and assumed that contact processes between individuals would follow Poisson statistics which can be described by an exponential distribution (\textit{i.e.,} homogeneity in the timing of events). However, an increasing number of empirical studies reveal that human temporal activity patterns are much more heterogeneous than the Poisson statistics considered, such as the burstiness ranging from the patterns of communication via emails~\cite{barabasi2005origin,eckmann2004entropy}, instant messaging~\cite{leskovec2008planetary}, web browsing~\cite{gonccalves2008human,radicchi2009human}, mobile phone calls~\cite{candia2008uncovering,karsai2011small} and text messages~\cite{wu2010evidence,zhao2011empirical}, to the patterns in physical contacts probed by wireless devices~\cite{cattuto2010dynamics,isella2011s} and sexual contact survey~\cite{rocha2010information,rocha2011simulated}. These temporal inhomogeneities can be well described by heavy-tailed or power-law distributions, which is in stark contradiction with the Poisson approximation.

The above temporal activities are characterized by the inter-event time (or waiting time) and response time, where the former is the time interval between two consecutive activities by the same user, and the later is the time interval between receiving a message or an email and replying to (or forwarding) it by one user. With the fact that spreading processes have to follow the time ordering of events, the temporal heterogeneities have a major impact on the spreading dynamics. In the past few years, researchers started to incorporate temporal heterogeneity into the spreading dynamics and studied its impacts with the help of Monte Carlo simulations and analytical tools. Vazquez \textit{et al.} first addressed the failure of the Poisson approximation for the inter-event times of human interactions and showed that the heterogeneous inter-event time distribution obviously slows down the spreading dynamics~\cite{vazquez2007impact}. Min \textit{et al.} provided a theoretical prediction to conclude that the heavy-tailed waiting time distribution leads to a power-law decay in the number of new infections in the long time limit, resulting in the extremely slow prevalence decay~\cite{min2010spreading}. They also showed that the heterogeneous waiting times in contact dynamics can observably impede epidemic spreading and the epidemic outbreak can even be completely suppressed when the temporal heterogeneity is strong enough~\cite{min2013absence}. They applied the renewal theory to analyze the Susceptible-Infected-Recovered (SIR) model and derived that the epidemic threshold increases with the heterogeneity of contact dynamics without bound. Karsai \textit{et al.} gave further insight into the impact of temporal heterogeneities on spreading dynamics and found that the slowing down of spreading is mainly caused by the weight-topology correlation and the heterogeneous inter-event times of individuals~\cite{karsai2011small}. Miritello \textit{et al.} indicated that bursts and group conversations have opposite effect on information spreading: bursts hinder the spreading at large scales, while group conversations make the spreading more efficient at local scales~\cite{miritello2011dynamical}. In contrast to these observations, Rocha \textit{et al.} found that the temporal correlations of inter-event times in sexual contact networks accelerate disease outbreaks~\cite{rocha2011simulated}. Furthermore, Takaguchi \textit{et al.} also reported that heterogeneous inter-event times facilitate epidemic spreading~\cite{takaguchi2012bursty}. From the above, the original cause of this difference still remains unclear~\cite{holme2011temporal,Masuda:2013b}.

Most of the recent studies focused on the impact of the heterogeneous inter-event or waiting times on spreading dynamics, while the influence of the heterogeneity of response times has attracted less attention. Iribarren and Moro performed a viral marketing experiment with emails and concluded that the large heterogeneity found in the response times (the time elapsed between receiving and forwarding) is responsible for the slow dynamics of information at the collective level~\cite{iribarren2009impact,iribarren2011branching}. However, the correlation between local structures and temporal activities was neglected in the above studies. Onnela \textit{et al.} empirically reported that in the mobile phone call network individuals who talk to a large number of friends appear to spend less time per friend than those who have few friends~\cite{onnela2007analysis}. Hidalgo \textit{et al.} showed that persistent links are more common for people with small degrees, while transient links are more common for individuals with large degrees, where persistent and transient links are differentiated by the quantity of persistence (defined as the probability that two neighbors at the ends of a link communicate with each other) on the link~\cite{hidalgo2008dynamics}. Given the limited time and energy, individuals with large degrees (\textit{e.g.,} contacting to a large number of friends in the email network) usually take a longer time to reply than those with small degrees~\cite{haerter2012communication}. In this study, we assume that the response time and the degree of each individual have positive correlation owning to the limitation of time and energy, and study the impact of heterogeneous response times on information spreading. The behavior of the Susceptible-Infective (SI) spreading dynamics incorporating the heterogeneous response times is investigated based on an uncorrelated scale-free network. The simulation results indicate that the heterogeneity of response times of individuals can speed up the spreading dynamics in the early and middle stages, that is, the stronger the heterogeneity of response times, the faster the spreading is, as this phenomenon is confirmed by the heterogeneous mean-field theory. For a given heterogeneity, we adjust the correlation between the response time and the degree of each individual and find that reducing the strength of correlation can also speed up the spreading dynamics in the early and middle stages. In the late stage, the full prevalence time no longer monotonously changes with the heterogeneity of response times and response time-degree correlation, but an optimal value occurs respectively.

The paper is organized as follows. In Sec.~\ref{SEC:model} , we introduce a SI model incorporating heterogeneous response times. In Sec.~\ref{SEC:3}, we study the effects of the heterogeneity of response times and the response time-degree correlation on the spreading dynamics in the early and middle stages. The impacts of the heterogeneity and the correlation on the spreading speed in the late stage are investigated in Sec. \ref{SEC:4}. Finally, we draw the conclusions in Sec. \ref{SEC:5}.

\section{Model Introduction}\label{SEC:model}

\subsection{Heterogeneous Response Times of Individuals}
Due to the limited time and capacity of each individual~\cite{haerter2012communication},
it takes a longer time to get the response from individuals who have many contacts,
compared to few contacts, in most cases. We therefore assume that the response time
$\tau_{i}$ of each individual $i$ replying to its neighbors is positively associated with
its degree $k_{i}$, as follows
\begin{equation}\label{taui}
\tau_{i}=A_{\alpha}(\frac{k_{i}}{k_{max}})^{\alpha},
\end{equation}
where $k_{max}$ is the maximum degree and used to normalize the response time~\cite{Yang:2008},
$A_{\alpha}$ and $\alpha$ are two tunable parameters.
The range of $\alpha$ is $\alpha\geq0$.
If $\alpha=0$, the response times of all the individuals are equal to $A_{\alpha}$.
If $\alpha > 0$, the response time $\tau_{i}$ for an individual with large degree $k_{i}$ takes longer.

To compare the cases for different values of $\alpha$, the value of $A_{\alpha}$ is to ensure that the mean response time over all responses on links $\langle \tau\rangle$ is equal~\cite{zhao2012epidemic,Yang:2011}, and the same total response time $T$ is thus
\begin{equation}\label{sumtau1}
T=\sum_{i=1}^{N}{\tau_{i}k_{i}},
\end{equation}
where $N$ is the network size.
Substituting Eq.~\eqref{taui} into Eq.~\eqref{sumtau1}, we have
\begin{equation}\label{sumtau2}
T=\sum_{i=1}^{N}A_{\alpha}(\frac{k_{i}}{k_{max}})^{\alpha}k_{i}.
\end{equation}
When $\alpha=0$, Eq.~\eqref{sumtau2} can be rewritten as
\begin{equation}\label{sumtau0}
\begin{aligned}
T=\sum_{i=1}^{N}A_{\alpha=0}k_{i}=A_{\alpha=0}\langle k\rangle N,
\end{aligned}
\end{equation}
where $\langle k\rangle$ is the mean degree of the network.
Combining Eqs.~\eqref{sumtau2} and \eqref{sumtau0}, we obtain
\begin{equation}\label{Aalpha}
A_{\alpha}=\frac{A_{\alpha=0}\langle k\rangle Nk_{max}^{\alpha}}{\sum_{i=1}^{N}k_{i}^{1+\alpha}}.
\end{equation}
Once the value of $A_{\alpha=0}$ is given, we can get the value of $A_{\alpha}$
according to Eq.~\eqref{Aalpha}.
Since degree and response time distributions are discrete,
we have $P(\tau_{k})=P(k)k$.
If the degree distribution follows a power law of $P(k)\sim k^{-\gamma}$ with $\gamma>1$,
then the response time distribution will be given by
\begin{equation}\label{Ptau}
P(\tau_{k})\sim \tau_{k}^{\frac{1-\gamma}{\alpha}}.
\end{equation}
The above equation implies that the distribution of response times
for large $\alpha$ is more heterogeneous.

\begin{figure}[h]
\centering
\includegraphics[width=8cm,height=6cm,angle=0]{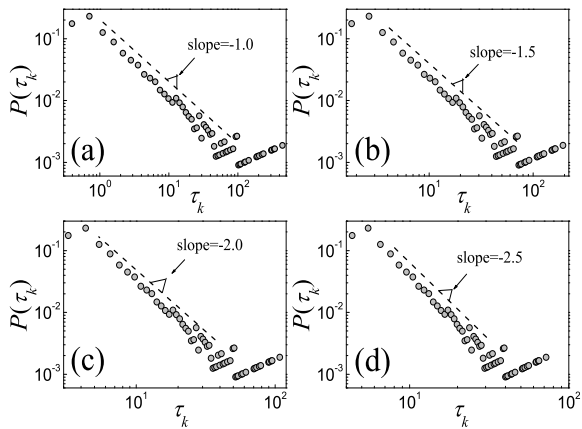}
\caption{Response time distributions with different exponents. Results for the distributions correspond to (a) $\alpha=2.0$, (b) $\alpha=1.3$, (c) $\alpha=1.0$, and (d) $\alpha=0.8$. Analytic treatment from Eq.~\eqref{Ptau} suggests a scaling behavior with exponent $-1.0$, $-1.5$, $-2.0$, and $-2.5$ for subfigures (a)-(d), as shown by the dash lines.}
\label{fig1}
\end{figure}

\subsection{Temporal SI Model}
For the epidemic spreading dynamics,
SI, SIS, or SIR models~\cite{anderson1992infectious} typically are studied
in homogeneous mixing environment or on static networks,
where the letters in the acronyms stand for to the different states
(Susceptible, Infected, or Recovered) of individuals
and their dynamics can be described by the changes of these states
due to the influence of others.
Owing to the simplicity of SI model,
the effects of heterogeneous response times on information spreading
can be easily understood.
Although the other models such as SIS are even more practical,
more parameters such as the recovery rate $\mu$
in the SIS model make the dynamics more complicated.
In the traditional SI model, each individual can only be in two states,
either susceptible or infected.
In the beginning, some individuals are selected to be
initially infected (\textit{i.e.}, seeds) and all other individuals are susceptible.
At each time step, each infected individual contacts all its neighbors,
and then the susceptible neighbors will be infected with probability $\beta$.
This process will continue until all the susceptible individuals reachable
from initial seeds are infected.

In the temporal network with heterogeneous response times,
each infected individual is no longer in contact with its neighbors at each time step.
For an infected individual $j$, it only contacts its neighbor $i$
at some discrete time steps with an invariable time interval $\tau_{i}$,
where $\tau_{i}$ is the response time of individual $i$ and obtained by Eq.~\eqref{taui}.
Individual $j$ firstly makes a request for a contact to its neighbor $i$ at step $t$.
After $\tau_{i}$ steps, individual $i$ responses to $j$ at step $t+(\tau_{i}+1)$,
and the contact occurs at the same step. If the individual $i$ is not infected,
the individual $j$ will make a request again without any delay,
ignoring the effect of waiting time distribution. In other words,
the individual $j$ contacts its neighbor $i$ at the time steps $t_{j0}+(\tau_{i}+1)$,
$t_{j0}+2(\tau_{i}+1)$, $t_{j0}+3(\tau_{i}+1)$, ..., rather than every time step,
where $t_{j0}$ is the time step at which the individual $j$ is infected.
This is the only difference from the traditional SI model.

In the spreading process, the infected density $\rho$ is
a key quantity to characterize the spreading speed.
At a fixed time $t$, the greater $\rho(t)$ value indicates the faster spreading.
Likewise, the spreading speed can also be characterized by the prevalence time $t(\rho)$
at which $\rho$ fraction of all nodes are infected, especially the full prevalence time
$t_f$~\cite{karsai2011small}.

\section{Spreading dynamics in the early and middle stages}\label{SEC:3}

\subsection{Effect of heterogeneous response times}
To study how the heterogeneous response times affect the speed of information spreading,
we perform extensive numerical simulations.
A network with scale-free degree distribution as an uncorrelated configuration model (UCM)
is first constructed by randomly choosing stubs and connecting them to form edges,
while avoiding multiple and self-connections~\cite{catanzaro2005generation}.
The UCM network is generated with size $N\!=\!10^{4}$, minimum degree $k_{min}\!=\!3$,
maximum degree $k_{max}=10^2$, and degree distribution $P(k)\sim k^{-\gamma}$ with
$\gamma\!=\!3$. And then, we calculate the response time of each individual before
simulating the SI spreading dynamics. The value of the parameter $A_{\alpha=0}$ in
Eq.~(\ref{sumtau0}) is set as $A_{\alpha=0}=10$. It's important to note that different
values of the parameter don't affect the simulation results qualitatively. For different
$\alpha$, the value of $A_{\alpha}$ is obtained by Eq.~\eqref{Aalpha}, and then the
response time of each individual is calculated according to Eq.~\eqref{taui}.
Fig.~\ref{fig1} shows the response time distributions for different
values of $\alpha$ on a log-log plot, and the distributions follow a power law with
adjustable exponent obtained by Eq.~\eqref{Ptau}. With the increase of $\alpha$, the
response times will display a more heterogeneous distribution.
To be specific, the nodes of small degrees have shorter response times,
while the hubs respond to a piece of information later.
It is mentioned that a response time $\tau_{i}$ is not an integer in general.
In numerical simulations, it is implemented in a probabilistic way.
Taking $\tau=3.6$ as an example, the response time of individual $i$ is either
$\tau_{i}=3$ with probability $0.4$ or $\tau_{i}=4$ with probability $0.6$.
Without any special statement, all the following simulation results will be
obtained by averaging over $10^{3}$ independent realizations,
based on the UCM scale-free network. Each realization starts with randomly selecting
a certain fraction $\rho(0)\!=\!0.005$ of the individual as infected seeds.
The infection rate is $\beta\!=\!0.1$. The simulations at different values of
$\rho(0)$ and $\beta$ will reveal the same conclusion.

\begin{figure}[h]
\includegraphics[width=9cm,height=7cm,angle=0]{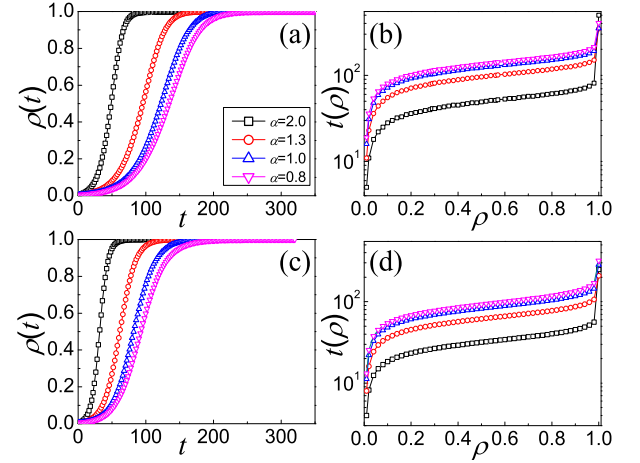}
\caption{The time evolution of information spreading with different response time distributions. The infected density $\rho(t)$ versus $t$ for simulated results (a) and theoretical predictions (c), and the prevalence time $t(\rho)$ versus $\rho$ for simulated results (b) and theoretical predictions (d). The results correspond to $\alpha=2$ (black squares), $\alpha=1.3$ (red circles), $\alpha=1$ (blue up triangles), and $\alpha=0.8$ (magenta down triangles), respectively.
}
\label{fig2}
\end{figure}

\begin{figure}[h]
\includegraphics[width=6cm,height=4.5cm,angle=0]{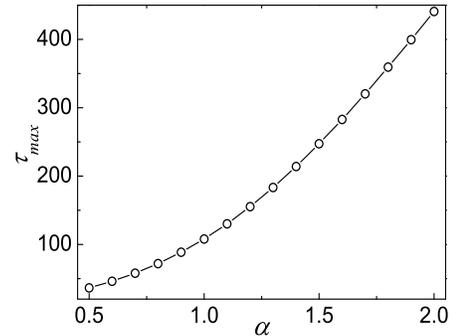}
\caption{The maximum response time $\tau_{max}$ as a function of $\alpha$.}
\label{fig3}
\end{figure}

Fig.~\ref{fig2} shows the time evolution of infected density $\rho(t)$ for different values of $\alpha$. The heterogeneity of the response times has a remarkable impact on information spreading, and the spreading dynamics can be accelerated in the early and middle stages of the spreading process by enhancing the heterogeneity (\textit{i.e.,} increasing the value of parameter $\alpha$), as shown in Fig.~\ref{fig2} (a). The phenomenon is more clearly indicated by Fig.~\ref{fig2} (b). The larger the value of $\alpha$ is, the less time steps are required to infect a certain fraction of the individuals in the early and middle stages (\textit{e.g.}, $\rho\leq0.98$). Owing to the same total response time for different values of $\alpha$, the larger the value of $\alpha$ is, the shorter response times most of the individuals with small degrees have (see Fig.~\ref{fig1}). Therefore, these individuals with small degrees will be infected much earlier and the spreading can be facilitated in the early and middle stages. However, the hubs with large degrees have longer response times $\tau_{max}$ in the case of more heterogeneous response time distribution. Fig.~\ref{fig3} shows that the maximum response time $\tau_{max}$ monotonously increases with $\alpha$. The spreading speed in the late stage (\textit{e.g.}, $\rho=1.0$) can thus be slightly more complicated, which is further studied in next section.

To further understand the above phenomenon qualitatively, the temporal network with heterogeneous response times is modeled by a weighted graph with the assumption that contact times are random, with a frequency proportional to the edge weight~\cite{holme2011temporal}. We thus provide a theoretical framework based on the mean-field rate equation~\cite{Barthelemy:2004}, which qualitatively captures the dynamics of the introduced SI model. Assuming that all the individuals with the same degree have the same probability of infection at any given time, we define $s_{k}(t)$ and $\rho_{k}(t)$ to be the probability that an individual with degree $k$ is susceptible and infected at time $t$, respectively. Obviously, the two variables obey the normalization condition:
\begin{equation}
s_{k}(t)+\rho_{k}(t)=1.
\end{equation}

If a susceptible individual $A$ with degree $k$ becomes infected, it has to contact the infection from one of its neighbors. Assuming that a particular neighbor $B$ of $A$ is in the infected state, it must have been infected by one of its remaining neighbors. Therefore, without the degree correlation, the density of infected neighbors of a node with degree $k$ is
\begin{equation}
\Phi_{k}(t)=\frac{\sum_{k^{\prime}}(k^{\prime}-1)P(k^{\prime})\rho_{k^{\prime}}(t)}{\langle k\rangle},
\end{equation}
where $\langle k\rangle=\sum_{k^{\prime}}k^{\prime}P(k^{\prime})$.

As the temporal network is modeled as a weighted graph, an individual of degree $k$ will contact with an infected neighbor with a frequency of $1/(\tau_k+1)$. Obviously, the larger $\tau_k$ indicates the less contact frequency between them, and vice versa. The evolution equation of $\rho_{k}(t)$ with degree $k$ is thus written as
\begin{equation}
\frac{d\rho_{k}(t)}{dt}=\frac{\beta k[1-\rho_{k}(t)]}{\tau_k+1}\Phi_{k}(t).
\end{equation}
When all $\tau_k$ values are equal to zero, the dynamics will return to the case of the classic SI model. At each time step, the infected density is given by
\begin{equation}
\rho(t)=\sum_{k}P(k)\rho_{k}(t).
\end{equation}
When all individuals are infected, \textit{i.e.,} $\rho(t_f)=1$, we also obtain the numerical solution of the full prevalence time $t_f$. From Figs.~\ref{fig2} (c) and (d), we see that the time evolutions of infected density and the prevalence time required to infect a certain fraction of the individuals, which are predicted by the above theoretical analysis, show a qualitative agreement with the observation of the simulated results. The delicate differences between the simulated results and the theoretical predictions are derived from the edge-weight simplification of temporal contacts between nodes~\cite{Masuda2013,Scholtes:2013}.

\subsection{Impact of Response Time-Degree Correlation}

In our model, the response times of each individual replying to all the neighbors are equal and proportional to its degree according to Eq.~\eqref{taui}.
In real life, however, we usually take a shorter time and make a priority response to some important people for us than unimportant ones within a time-limited environment. Haerter \textit{et al.}~\cite{haerter2012communication} reported that a user preferentially replies to the senders with large degrees on finite capacity social networks. From this we know that an individual may take different response times to reply to its neighbors, that is to say that the response times are not always perfectly correlated with nodes' degrees. Here we investigate how the correlation between the response times and degrees of individuals affects the spreading dynamics. To be concrete, we focus on the effect of positive correlation on the spreading speed.

The strength of the correlation is measured by the Pearson correlation coefficient $r$~\cite{newman2002assortative}, which is computed according to all data element pairs $(k_{i}, \tau_{ij})$, where $k_{i}$ is the degree of individual $i~(i=1,2,...,N)$, and $\tau_{ij}$ is the response time of individual $i$ replying to its neighbor $j~(j=1,2,...,k_{i})$. It lies in the range $[-1,1]$ in the thermodynamic limit, where the special case of $r=0$ is achieved in the case of no correlation. A temporal network with adjustable strength of response time-degree correlation can be realized by reducing the correlation between response times and degrees of individuals on the original temporal network. Once a scale-free network is given, the response time of each individual replying to its neighbors is obtained by Eq.~\eqref{taui}, on which there is a maximally positive correlation $r_{max}$. We adjust the correlation between the response times and degrees of individuals using a similar algorithm in Ref.~\cite{xulvi2004reshuffling}. The step of our algorithm looks as follows. At each step, two random individuals of the network are chosen, and then a response time for each replying to one of its neighbors is randomly selected, so that we could consider the two response times of two individuals being with different degrees in general, which can be exchanged in two situations: when the degree and response time of an individual are both greater (smaller) than the other's. After each exchange, we recalculate the Pearson correlation coefficient $r$. This process will continue until $r$ is equal to or less than a set value.

\begin{figure}[h]
\includegraphics[width=8cm,height=6cm,angle=0]{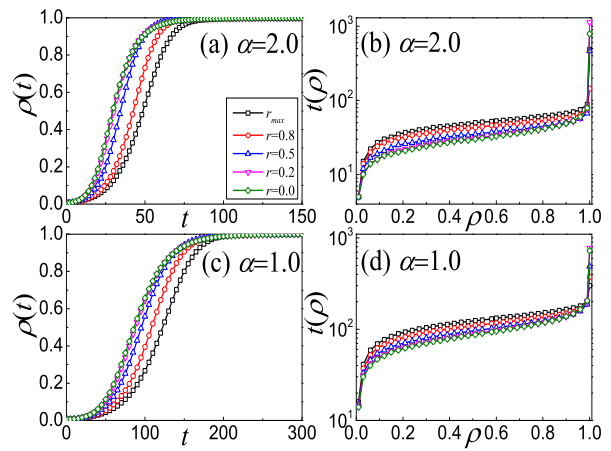}
\caption{The time evolution of SI spreading dynamics with different response time-degree correlations. The infected density $\rho(t)$ versus $t$ at $\alpha=2.0$ (a) and $\alpha=1.0$ (c), and the prevalence time $t(\rho)$ versus $\rho$ at $\alpha=2.0$ (b) and $\alpha=1.0$ (d). In each subfigure, five different values of Pearson correlation coefficient $r$ are chosen ($r_{max}, 0.8, 0.5, 0.2$, and $0.0$), corresponding to the black squares, red circles, blue up triangles, magenta down triangles, and green diamonds, respectively.
}
\label{fig4}
\end{figure}

To investigate the impact of the response time-degree correlation on the spreading dynamics, Monte Carlo simulations are also performed. We first adjust the correlation according to the discussed algorithm and then carry out the spreading dynamics. For a given heterogeneity (\textit{i.e.,} $\alpha=2.0,1.0$), Fig.~\ref{fig4} shows the infected density $\rho(t)$ and the prevalence time $t(\rho)$ at different values of $r$. In the early and middle stages, we see that the lower the correlation is, the faster the spreading dynamics is. As the nodes with large degrees may have some shorter response times at a smaller value of $r$, the hubs will be infected earlier, which can facilitate information spreading. This phenomenon is also indicated by Fig.~\ref{fig5} (b). In addition, we also see that the change of the prevalence time $t(\rho)$ for $\rho=0.9$ is very small when $r\leq0.6$, which means too much shuffling (\textit{i.e.,} randomly exchanged response times) is superfluous for improving the spreading speed. With the decrease of $r$, some individuals with small degrees may have longer response times. Thus, it needs a few more steps to infect the individuals with small degrees but longer response times, which may make the spreading speed in the late stage display a complete opposite trend to that in the early and middle stages.

\section{Spreading dynamics in the late stage}\label{SEC:4}
\begin{figure}[h]
\includegraphics[width=8cm,height=6cm,angle=0]{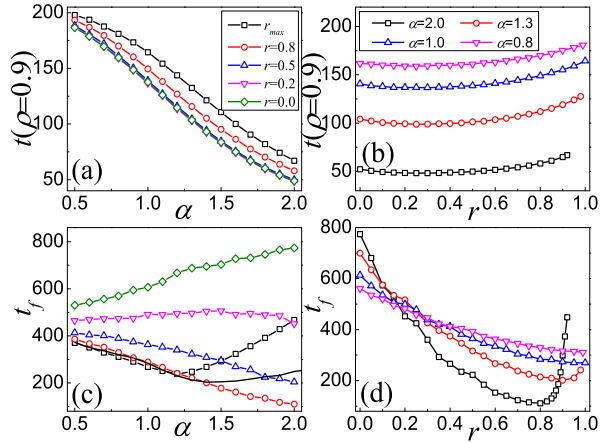}
\caption{The prevalence time under different parameters. For $\rho=0.9$, $t(\rho=0.9)$ versus $\alpha$ (a) and $r$ (b), and $t_f$ versus $\alpha$ (c) and $r$ (d) for $\rho=1.0$. In subfigures (a) and (c), the results at $r_{max}, 0.8, 0.5, 0.2$, and $0.0$ are denoted by the black squares, red circles, blue up triangles, magenta down triangles, and green diamonds, respectively. In subfigures (b) and (d), the results at $\alpha=2.0, 1.3, 1.0$, and $0.8$ are denoted by the black squares, red circles, blue up triangles, and magenta down triangles, respectively. Note that the values of $r_{max}$ in the original networks with $\alpha=2.0, 1.3, 1.0, 0.8$ are $r_{max}=0.92, 0.98, 1.00, 0.99$, respectively, on account of finite size effect. The black solid line in subfigure (c) comes from the analytical predictions of the heterogeneous mean-field theory.
}
\label{fig5}
\end{figure}

\begin{figure}[h]
\includegraphics[width=8cm,height=6cm,angle=0]{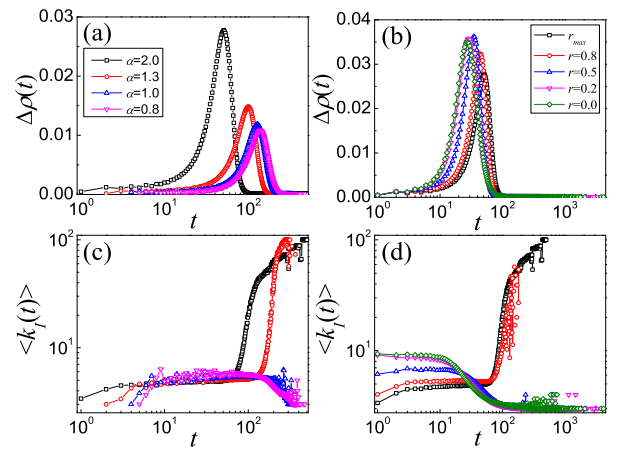}
\caption{The time behavior of the newly infected density $\Delta \rho(t)$ (a) and (b), and the average degree over the newly infected individuals $\langle k_{I}(t)\rangle$ (c) and (d). (a) and (c) show the simulated results on the original temporal networks with $r_{max}$ at different values of $\alpha$, (b) and (d) show the results for different values of $r$ when $\alpha=2.0$.}
\label{fig6}
\end{figure}

In this section, we focus on the influence of the heterogeneity and the correlation on the spreading dynamics in the late stage. The full prevalence time $t_f$ as a function of the heterogeneity parameter $\alpha$ and the correlation coefficient $r$ is shown in Fig.~\ref{fig5}. On the original temporal network with $r_{max}$, Fig.~\ref{fig5} (a) shows that the more heterogeneous (corresponding to the greater value of $\alpha$) the response time is, the less prevalence time for $\rho=0.9$ is required. While it is not the case for the fully prevalence time $t_f$, and there is an optimal value at $\alpha \approx 1.2$ with the least time steps to infect all the individuals in Fig.~\ref{fig5} (c). To find out why, the time evolutions of the newly infected density $\Delta \rho(t)$ and the average degree over the newly infected individuals $\langle k_{I}(t)\rangle$ are shown in Figs.~\ref{fig6} (a) and (c), respectively. From Fig.~\ref{fig6} (c), we see that the individuals with small degrees are infected in the late stage when $\alpha \textless 1.2$. As the response times of individuals with small degrees decrease with the increase of $\alpha$, these individuals will be infected much earlier and the full prevalence time will be shortened. However, when $\alpha$ increases to a certain value and continues (\textit{i.e.,} $\alpha \textgreater 1.2$), $\tau_{max}$ is getting longer and longer. In this case, the hubs with $\tau_{max}$ become a key factor to hold back the spreading dynamics, which increases the full prevalence time. For $\alpha=2.0$, $90\%$ nodes are infected in $t\approx70$ time steps (see Fig.~\ref{fig5} (a)), while the hubs are infected after $\tau_{max}\approx450$ time steps at least (see Fig.~3). It is also indicated by $\langle k_{I}(t)\rangle$ in the late stage for $\alpha=1.3$ and $2.0$ as shown in Fig.~\ref{fig6} (c).

With the decrease of $r$, the full prevalence time $t_f$ displays a distinctly different trend as shown in Fig.~~\ref{fig5} (c). $t_f$ monotonically decreases with $\alpha$ when $r=0.5$ and $r=0.8$, while $t_f$ monotonically increases with $\alpha$ when $r=0.0$. For $r=0.8$ and $r=0.5$, some randomly exchanged response times on links make the individuals with large degrees have shorter response times, and these individuals will be infected earlier. From Fig.~~\ref{fig6} (d), we see when $\alpha=2.0$, $\langle k_{I}(t)\rangle$ for $r=0.8$ is smaller than that for $r_{max}$ in the late stage. Therefore, the inhibition of these individuals is subsided and $t_f$ displays a relatively consistent trend with that in the early and middle stages. For $r=0.2$, the heterogeneity of response times has little influence on the full prevalence time (\textit{i.e.,} almost the same for different values of $\alpha$), which originates from that the facilitation of individuals with small degrees falls off rapidly and almost disappears. For $r=0.0$, the full prevalence time increases with the heterogeneity of response times. Considering all response times on links are randomly distributed in the absence of correlation, the maximum response time $\tau_{max}$ is more likely to be allocated to an individual with small degree. As the temporal network with more heterogeneous response time distribution (\textit{i.e.,} greater $\alpha$) has a longer $\tau_{max}$, the individuals with small degrees but $\tau_{max}$ will be infected later, which slows down the spreading dynamics.

Figs.~\ref{fig5} (b) and (d) report that the prevalence times for $\rho=0.9$ and $\rho=1.0$ change with the Pearson correlation coefficient $r$ at a given heterogeneity. In Fig.~\ref{fig5} (b), the less prevalence time $t(\rho)$ for $\rho=0.9$ is shown relatively to the decrease of $r$, which means randomly shuffling can speed up the spreading in the early and middle stages. In Fig.~5 (d), there is an intersection at about $r=0.2$, which further confirms the results for $r=0.2$ in Fig.~5 (c). When $\alpha=2.0$, an optimal value occurs at $r \approx 0.8$ with the least full prevalence time. From Fig.~\ref{fig6} (d), we see that the average degree over the newly infected individuals $\langle k_{I}(t)\rangle$ increases with time $t$ at a high correlation (\textit{e.g.,} $r_{max}=0.92\geq0.8$), while decreases with time $t$ at a low correlation (\textit{e.g.,} $r=0.2<0.8$). For $r \geq 0.8$, the response times of individuals with large degrees decrease with the decrease of $r$, which is beneficial to the spreading and results in the full prevalence time shortened. Well, when $r$ reduces to some extent and continues (\textit{e.g.,} $r\textless 0.8$ for $\alpha=2.0$), the individuals with small degrees but longer response times strongly hinder the spreading dynamics, which increases the full prevalence time. When $\alpha=1.0$ and $\alpha=0.8$, $t_f$ increases with the decrease of $r$. It is because that the $\tau_{max}$ of hubs doesn't hold back the spreading dynamics when $\alpha<1.2$, while the random exchange of response times only makes the average response times of individuals with small degrees increase. The experiments shown here illustrate that the heterogeneity of response times of individuals and the correlation between the response times and degrees of individuals have a significant impact on the spreading dynamics, which not only contributes to further understanding the effect of human temporal activity patterns on the information spreading, but also provides potential measures to control or speed up the information spreading.

\section{Conclusion And Discussion}\label{SEC:5}
In conclusion, we have introduced a temporal model to study the effect of the heterogeneous response times on information spreading.
We assume that the response time $\tau$ of each individual is positively associated with its degree $k$, and the power-law response time distribution with adjustable exponent is obtained on the uncorrelated scale-free network. Numerical study shows that the spreading dynamics can be accelerated by enhancing the heterogeneity of the response time distribution in the early and middle stages, which is derived from that the individuals with small degrees have shorter response times for the larger value of $\alpha$. This phenomenon can be further validated by the mean-field theoretical analysis. The more heterogeneous the response time is, the faster the spreading is in the early and middle stages. To be more realistic, we have also studied the impact of the response time-degree correlation measured by the Pearson correlation coefficient $r$ on the spreading dynamics, and found that for a given heterogeneity reducing the strength of correlation can speed up the spreading in the early and middle stages.

However, in the late stage, the full prevalence time no longer monotonously changes with the increase of $\alpha$, but an optimal value occurs. In other words, increasing the heterogeneity of response times doesn't always accelerate the spreading in the late stage. And more notably, there is also an optimal value of the full prevalence time changing with $r$. These two optimal values both stem from the reasonable allocation of heterogeneous response times, which ensures that not only the individuals with large degrees have shorter response times, but also the individuals with small degrees are infected in the early or middle stages.

The above results are not reported by previous researches, which is conductive to our understanding and optimizing information spreading. To summarize, this work provides complementary information to the previous studies and contributes to understanding the impact of the heterogeneous human activities on information spreading, such as the forwarding of emails and the invasion of mobile phone viruses. The study of human activity patterns and their effects on various dynamics is still a relatively new field with many open questions~\cite{holme2011temporal,Masuda:2013b,HolmeP:2013c}, such as the phase transition phenomena of the SIS and SIR model. In addition, an exact analytic method is required urgently. As the heterogeneous mean-field method neglects the non-Poissonian time property and the strong dynamical correlation, it only qualitatively captures the dynamics. Although the number of infected individuals can be worked out quantificationally by branching process, it will fail once the population is finite or the infection probability is large~\cite{iribarren2011branching}. A more suitable theoretical method of the temporal SI model still needs to think deeply. For instance, in the case of non-Poissonian inter-event time distributions on the SI spreading dynamics in homogeneous mixing environment~\cite{JoHH:2013}, bursty dynamics can accelerate the spreading for early and intermediate times, while the behavior is opposite for late time dynamics.

\acknowledgments
This work was partially supported by the NNSF of China (Grant Nos. 11105025, 61103109, 91324002, 11305078) and the Fundamental Research Funds for the Central Universities (Grant No. ZYGX2012YB027). Y. Do was supported by Basic Science Research Program through the National Research Foundation of Korea (NRF) funded by the Ministry of Education, Science and Technology (NRF-2013R1A1A2010067).

\nocite{*}
\bibliographystyle{apsrev}
\bibliography{reference}

\end{document}